# Implementation of quantum teleportation protocol by superposition of displacement operators with absolute equal but opposite in sign amplitudes

Sergey A. Podoshvedov

Department of computer modeling and nanotechnology, Institute of natural and exact sciences, South Ural State University, Lenin Av. 76, Chelyabinsk, Russia
e-mail: sapodo68@gmail.com

**Abstract** We consider novel implementation of quantum teleportation protocol of unknown qubit. Entangled hybrid state with coherent components of small amplitude is used as quantum channel. Action of the channel on the teleported qubit can be approximated by superposition of displacement operators with equal modulo but opposite in sign amplitudes. Coherent components of the quantum channel displace the teleported qubit at the same absolute but opposite in sign values, so that any information about amount by which the qubit is displaced is lost despite the fact that one of these events has definitely occurred. Alice unambiguously distinguishes her measurement outcomes and Bob obtains at his disposal two states of a single photon in superposition state of two modes with controllable amplitude distortions. 2 bits of classical information is required for Bob to identify the states. Initial driven amplitude modulation of the unknown qubit enables to increase efficiency of the protocol and implement it in nearly deterministic manner for highly unbalanced teleported qubit. Remarkably, the success probability of the teleportation can be made arbitrary high for amplitude modulated unknown qubit taken as original. Amplitude modulation of unknown qubit is nearly deterministic procedure. The scheme is implemented with irreducible number of optical elements: one beam splitter and two on-off detectors. We show the quantum channel is realized in the same manner and its implementation requires one additional beam splitter.

## 1. Introduction

Creation of a quantum computer capable of realizing algorithms such as quantum factoring [1] and quantum search [2] require both a design a universal set of gate operations for a large system and good fault-tolerant procedures to overcome inevitable imprecisions in unitary evolution of the physical system. There are many suggested approaches for quantum computers, but none of them are completely satisfactory, in the sense, the proposed methods are quite complex and can require an unacceptable number of additional operations to produce a specific desired reversible operations [3-5]. So, the question of resources (mechanisms, approaches, states) needed to realize scalable quantum computing is currently still open.

Light states are good candidates for quantum information processing [6]. So, single-qubit operations with photon states can be directly realized by linear optics methods. Immediate difficulties arise in the implementation of entangling gates with photon states as it is difficult to make photons interact with each other in a desired manner. The standard idea to implement entangling gates with optical states in practice is based on the teleportation protocol [7] and Bell-state measurement with linear optics [8]. The success probability of the Bell-state measurement with linear optics elements and photodetectors does not exceed $0.5$ [8-10]. Controlled operations like to controlled$-X$ operation can be performed by simultaneous teleportation of two arbitrary qubits through entangled quantum channel [4], therefore success probability of such gates is limited to $0.25$ [11-13].

The other line of quantum information processing by optical qubits has been devoted to implementation with continuous variable states whose observable has a continuum of eigenvalues [14]. In the encoding, two coherent states $\left|-\alpha\right\rangle$ and $\left|\alpha\right\rangle$ are used as base elements. The Bell-state measurement for coherent qubits can be performed in a nearly deterministic manner with $\alpha$ growing provided that single-qubit operations can be realized with the coherent states [15]. Entangled coherent states can be discriminated through photon number resolving detection (PNRD) [16] that is not easy to implement in practice. In general, approaches with discrete variable states can achieve fidelity close to unity but at the expense of the efficiency of processes (probabilistic restrictions), while continuous variable states suffer from strong sensitivity to loses and inevitable limited fidelities. Idea to combine the two approaches and use their best properties looks natural. The idea of hybridization between discrete variable and continuous variable states [17] can be exploited to have serious advantages in realization of quantum protocols and quantum computation [18-21]. Recently, some implementations of hybrid entanglement between a coherent qubit (superposition of coherent states (SCS)) and microscopic qubit of vacuum and single photon [22] and a single photon in polarization basis [23] were demonstrated.

Here, we develop novel way to implement quantum teleportation protocol. Hybrid entanglement is used as quantum channel for transmission of quantum information from superposition state of vacuum and single photon to the state of single photon occupying two modes. Driven force for the teleportation can be approximated by superposition of the displacement operators with opposite in sign amplitudes. The approach does not use Bell state formalism [24]. Coherent components of hybrid channel simultaneously displace unknown teleported qubit in indistinguishable manner on a highly transmissive beam splitter (HTBS) by the values that differ from each other only by sign $\pm\alpha$. Given operation is unconditional. Choice of the displacement amplitude $\alpha << 1$ greatly simplifies implementation of the protocol of quantum teleportation and removes the requirement of PNRD. The teleported state and coherent part of hybrid channel disappear at the place of the measurement [7] (registration of some events in coherent and teleported modes) and it projects the single photon located arbitrary far from away into one of two possible states than can be unambiguously identified by the receiving party using 2 bits of classical information dispatched by the sender. Bob's two states undergo amplitude distortion. But Alice has a wide choice to manipulate the initial unknown qubit (controllable amplitude modulation) in order to increase the success probability of the protocol, which can reach almost unit value. Mathematical apparatus for the development of the protocol is based on representation of the displaced number states in terms of the number states [21,25]. This method has been used to generate even and odd SCS states of large amplitude by subtraction of photons from squeezed coherent state regardless of the number of subtracted photons [26,27] as well as to consider feasibility of one-dimensional rotations of coherent states (Hadamard gate) [28].

A detailed review of the displacement operations [29,30] is presented in section 2. In section 3, we describe direct implementation of the quantum teleportation protocol. We show its strengths and discuss the problem of amplitude distortion. In section 4, we discuss the methods of increase of its efficiency by controllable amplitude modulation of unknown qubit. In section 5, we show applicability of the mechanism to generate hybrid channel and nearly deterministically implement controllable amplitude modulation. Section 6 generalizes key moments of the studied quantum teleportation protocol. Additional auxiliary mathematical apparatus with displaced states of photons is presented in Appendix A.

## 2 Realization of displacement operators

Before considering the protocol of quantum teleportation of unknown qubit through a hybrid channel, let us consider interaction of strong coherent field with arbitrary state on HTBS

$$BS = \begin{vmatrix} t & -r \\ r & t \end{vmatrix}, \tag{1}$$

where $t$, $r$ are the transmittance $t \to 1$ and reflectance $r \to 0$, respectively, satisfying the normalization condition $t^2 + r^2 = 1$. We consider that the parameters $t$ and $r$ are the real values. So, interaction of two coherent states $|\alpha\rangle_1|\beta\rangle_2$ with amplitudes $\alpha$ and $\beta$, respectively, gives outcome

$$BS|\alpha\rangle_1|\beta\rangle_2 = |\alpha t + \beta r\rangle_1|\beta t - \alpha r\rangle_2. \tag{2}$$

where subscripts denote the state modes. Let $\rho_1$ be an arbitrary state which can be written in terms of the Glauber – Sudarshan $P$ – function as

$$\rho_1 = \int d^2\alpha P(\alpha)|\alpha\rangle_1\langle\alpha|. \tag{3}$$

Consider interaction of the state with coherent state on HTBS (1). By virtue of (2), we have

$$BS\left(\rho_1 \otimes |\beta\rangle_2\langle\beta|\right)BS^+ = \int d^2\alpha P|\alpha t + \beta r\rangle_1\langle\alpha t + \beta r| \otimes |\beta t - \alpha r\rangle_2\langle\beta t - \alpha r|. \tag{4}$$

where the notation $\otimes$ means tensor product. The integral (4) can be transformed to

$$\int d^2\alpha P|\alpha + \gamma\rangle_1\langle\alpha + \gamma| \otimes |\beta\rangle_2\langle\beta| = \int d^2\alpha P D(\gamma)|\alpha\rangle_1\langle\alpha|D^+(\gamma) \otimes |\beta\rangle_2\langle\beta|. \tag{5}$$

in the limit case of [31]

$$t \to 1,\ r \to 0, \text{ but } \beta r \to \gamma, \tag{6}$$

where the displacement operator (A1) with $\gamma$ being the displacement amplitude is used. Then finally, we can rewrite (4)

$$BS\left(\rho_1 \otimes |\beta\rangle_2\langle\beta|\right)BS^+ \approx D_1(\gamma)\rho_1 D_1^+(\gamma) \otimes |\beta\rangle_2\langle\beta|, \tag{7}$$

which in the case of pure state $\rho_1 = |\Psi\rangle_1\langle\Psi|$ implies

$$BS\left(|\Psi\rangle_1 \otimes |\beta\rangle_2\right) \approx D_1(\gamma)|\Psi\rangle_1 \otimes |\beta\rangle_2. \tag{8}$$

The condition (6) means that amplitude of auxiliary coherent state $|\beta\rangle_2$ tends to infinity $\beta \to \infty$ [31]. In real experiment, the amplitude of the coherent state to implement deterministic displacement of arbitrary state must be chosen to be sufficiently large. If we apply the coherent state with negative amplitude $|-\beta\rangle$, then the result of interaction on HTBS can be approximated by

$$BS\left(|\Psi\rangle_1 \otimes |-\beta\rangle_2\right) \approx D_1(-\gamma)|\Psi\rangle_1 \otimes |-\beta\rangle_2. \tag{9}$$

Expressions (8,9) are applicable to consideration of the teleportation protocol shown in Fig. 1. Suppose Alice wants to teleport unknown qubit

$$|\varphi\rangle_2 = a_0|0\rangle_2 + a_1|1\rangle_2, \tag{10}$$

to Bob which is located at a considerable distance apart from Alice. Alice cannot send this qubit directly but she has at her disposal a part of quantum channel,

$$|\Psi\rangle_{134} = \left(|0,-\beta\rangle_1|01\rangle_{34} + |0,\beta\rangle_1|10\rangle_{34}\right)/\sqrt{2}, \tag{11}$$

which is created in advance and connects Alice and Bob. Here, the notation for displaced number states (A2) is used. The amplitude of the state $\beta$ is assumed to be positive $\beta > 0$ throughout the consideration. The quantum channel is an entangled hybrid state, which consists of coherent components belonging to Alice (mode $1$) and one photon (dual-rail photon), which simultaneously take two modes (modes $3$ and $4$) at Bob's disposal. Alice mixes unknown qubit (10) with her coherent components on HTBS as shown in Fig. 1. The outcome of the mixing is given by

$$BS_{12}\left(|\Psi\rangle_{134}|\varphi\rangle_2\right) = \left(BS_{12}\left(|0,-\beta\rangle_1|\varphi\rangle_2\right)|01\rangle_{34} + BS_{12}\left(|0,\beta\rangle_1|\varphi\rangle_2\right)|10\rangle_{34}\right)/\sqrt{2}, \tag{12}$$

due to linearity of the unitary beam splitter operator (1). Let us consider the action of coherent components on the teleported state separately. So, outcome of the mixing of the coherent state $|0,-\beta\rangle$ with the unknown qubit (10) results in

$$\begin{aligned}&BS_{12}\left(\left|0,-\beta\right\rangle_1\left|\varphi\right\rangle_2\right)\left|01\right\rangle_{34}=BS_{12}\left(D_1(\beta)D_2(-\alpha)\right)BS_{12}^{+}BS_{12}\left(\left|0\right\rangle_1D_2(\alpha)\left|\varphi\right\rangle_2\right)\left|01\right\rangle_{34}=\\&D_1(-\beta/t)D_2(0)BS_{12}\left(\left|0\right\rangle_1\left(a_0\left|0,\alpha\right\rangle_2+a_1\left|1,\alpha\right\rangle_2\right)\right)\left|01\right\rangle_{34}=\\&FD_1(-\beta/t)BS_{12}\left(\left|0\right\rangle_1\sum_{m=0}^{\infty}\left(a_0c_{0m}(\alpha)+a_1c_{1m}(\alpha)\right)\left|m\right\rangle_2\right)\left|01\right\rangle_{34}=\\&F\left|0,-\beta/t\right\rangle_1\left(\sum_{m=0}^{\infty}t^m\left(a_0c_{0m}(\alpha)+a_1c_{1m}(\alpha)\right)\left|m\right\rangle_2\right)\left|01\right\rangle_{34}+\\&FD_1(-\beta/t)\left(\sum_{m=0}^{\infty}\left(a_0c_{0m}(\alpha)+a_1c_{1m}(\alpha)\right)\left|X^{(m)}\right\rangle_{12}\right)\left|01\right\rangle_{34}\end{aligned} \quad . \quad (13)$$

Using the same calculation technique, one obtains output state being result of interaction of the coherent state $\left|0,\beta\right\rangle$ and the teleported state (10) as inputs to the HTBS in Fig. 1

$$\begin{aligned}&BS_{12}\left(\left|0,\beta\right\rangle_1\left|\varphi\right\rangle_2\right)\left|10\right\rangle_{34}=BS_{12}\left(D_1(\beta)D_2(\alpha)\right)BS_{12}^{+}BS_{12}\left(\left|0\right\rangle_1D_2(-\alpha)\left|\varphi\right\rangle_2\right)\left|10\right\rangle_{34}=\\&D_1(\beta/t)D_2(0)BS_{12}\left(\left|0\right\rangle_1\left(a_0\left|0,-\alpha\right\rangle_2+a_1\left|1,-\alpha\right\rangle_2\right)\right)\left|10\right\rangle_{34}=\\&FD_1(\beta/t)BS_{12}\left(\left|0\right\rangle_1\sum_{m=0}^{\infty}\left(a_0c_{0m}(-\alpha)+a_1c_{1m}-(\alpha)\right)\left|m\right\rangle_2\right)\left|10\right\rangle_{34}=\\&F\left|0,\beta/t\right\rangle_1\left(\sum_{m=0}^{\infty}(-1)^mt^m\left(a_0c_{0m}(\alpha)-a_1c_{1m}(\alpha)\right)\left|m\right\rangle_2\right)\left|10\right\rangle_{34}+\\&FD_1(\beta/t)\left(\sum_{m=0}^{\infty}(-1)^m\left(a_0c_{0m}(\alpha)-a_1c_{1m}(\alpha)\right)\left|X^{(m)}\right\rangle_{12}\right)\left|10\right\rangle_{34}\end{aligned} \quad . \quad (14)$$

Here, we made use of unitary properties of the beam splitter and displacement operators $BS(BS)^{+}=I$ and $D(\alpha)(D(\alpha))^{+}=D(\alpha)D(-\alpha)=I$, respectively, where $I$ is an identity operator and notation + means Hermitian conjugate [34]. The displacement amplitude $\alpha$ is chosen to fulfill $\alpha=\beta r/t$. The decompositions of the coherent and displaced single photon states [21] over the number states (A9,A10) are used in derivation of (13,14). The change in the sign of the matrix elements under change of the displacement amplitude on opposite in sing (A13-A17) is a key moment for generating an output entangled states composed of quantum channel (10) and teleported state (11). Expressions (13,14) involve output state being result of passing the number state $\left|m\right\rangle$ through the HTBS [34]

$$BS\left(\left|0\right\rangle_1\left|m\right\rangle_2\right)=\left(\left(ra_1^{+}+ta_2^{+}\right)^m/\sqrt{m!}\right)\left|00\right\rangle_{12}=\left|X^{(m)}\right\rangle_{12}+t^m\left|0\right\rangle_1\left|m\right\rangle_2, \quad (15)$$

where

$$\left|X^{(m)}\right\rangle_{12}=\left(1/\sqrt{m!}\right)\sum_{k=0}^{m-1}\frac{r^{m-k}t^km!}{k!(m-k)!}\sqrt{k!(m-k)!}\left|m-k\right\rangle_1\left|k\right\rangle_2 . \quad (16)$$

Summing up the formulas (13,14), one obtains the final state

$$BS\left(\left|\Psi\right\rangle_{1234}\left|\varphi\right\rangle_2\right)=\left|\Delta_1\right\rangle_{1234}+\left|\Delta_2\right\rangle_{1234}, \quad (17)$$

where

$$\left|\Delta_1\right\rangle_{1234}=\frac{F}{\sqrt{2}}\begin{pmatrix}\left|0,-\beta/t\right\rangle_1\left(\sum_{m=0}^{\infty}t^m\left(a_0c_{0m}(\alpha)+a_1c_{1m}(\alpha)\right)\left|m\right\rangle_2\right)\left|01\right\rangle_{34}+\\\left|0,\beta/t\right\rangle_1\left(\sum_{m=0}^{\infty}(-1)^mt^m\left(a_0c_{0m}(\alpha)-a_1c_{1m}(\alpha)\right)\left|m\right\rangle_2\right)\left|10\right\rangle_{34}\end{pmatrix}, \quad (18)$$

$$\left|\Delta_2\right\rangle_{1234} = \frac{F}{\sqrt{2}} \begin{pmatrix} D_1(-\beta/t)\left(\sum_{m=0}^{\infty}\left(a_0 c_{0m}(\alpha) + a_1 c_{1m}(\alpha)\right)\left|X^{(m)}\right\rangle_{12}\right)\left|01\right\rangle_{34} + \\ D_1(\beta/t)\left(\sum_{m=0}^{\infty}(-1)^m\left(a_0 c_{0m}(\alpha) - a_1 c_{1m}(\alpha)\right)\left|X^{(m)}\right\rangle_{12}\right)\left|10\right\rangle_{34} \end{pmatrix}. \quad (19)$$

The contribution of first term (18) in sum (17) prevails over the second (19) in the case of $t \to 1$, $r \to 0$. If $t \to 1$ and $r \to 0$ , then the final state (17) tends to

$$BS_{12}\left(\left|\Psi\right\rangle_{134}\left|\varphi\right\rangle_2\right) \to \left|\Delta_1^{(id)}\right\rangle_{1234}, \quad (20)$$

where the ideal normalized state becomes

$$\left|\Delta_1^{(id)}\right\rangle_{1234} = \frac{F}{\sqrt{2}} \begin{pmatrix} \left|0,-\beta\right\rangle_1\left(\sum_{m=0}^{\infty}\left(a_0 c_{0m}(\alpha) + a_1 c_{1m}(\alpha)\right)\left|m\right\rangle_2\right)\left|01\right\rangle_{34} + \\ \left|0,\beta\right\rangle_1\left(\sum_{m=0}^{\infty}(-1)^m\left(a_0 c_{0m}(\alpha) - a_1 c_{1m}(\alpha)\right)\left|m\right\rangle_2\right)\left|10\right\rangle_{34} \end{pmatrix}. \quad (21)$$

where now amplitude $\beta$ of the quantum channel (11) and the displacement amplitude $\alpha$ are connected by

$$\alpha = \beta\sqrt{1-t^2}\ . \quad (22)$$

In real case of HTBS with non-zero reflectance, the state (21) can only approximate outcome. By analogy with (8,9), we can present final state as

$$BS_{12}\left(\left|\Psi\right\rangle_{134}\left|\varphi\right\rangle_2\right) \approx \left|\Delta_1^{(id)}\right\rangle_{1234}. \quad (23)$$

The fidelity of such approximation can be evaluated by

$$Fid = \left|\left\langle\Delta_1^{(id)}\right|\left(\left|\Delta_1\right\rangle + \left|\Delta_2\right\rangle\right)\right|^2. \quad (24)$$

Unit fidelity means the states are identical to each other [24]. Substituting the considered states into (24), one obtains analytical expression for the fidelity

$$Fid = F^4\,\frac{\exp\left(-|\beta|^2(1-1/t)^2\right)}{4}\left(\sum_{m=0}^{\infty} t^m\left|f_m(-\alpha)\right|^2 + \sum_{m=0}^{\infty} t^m\left|f_m(\alpha)\right|^2\right)^2, \quad (25)$$

where

$$f_m(\pm\alpha) = a_0 c_{0m}(\pm\alpha) + a_1 c_{1m}(\pm\alpha). \quad (26)$$

Here, we neglect the terms proportional to $\sim r^2$ whose contribution is substantially insignificant in the case of $r << 1$. The fidelity becomes $Fid = 1$ in the case of $t = 1$ due to normalization conditions (A20). Now, we can talk about approximation of the interaction of the quantum channel (11) with the teleported state (10) by the following operator

$$\Omega = \left(\left|0,-\beta\right\rangle_1 D_2(\alpha)\left|01\right\rangle_{34} + \left|0,\beta\right\rangle_1 D_2(-\alpha)\left|10\right\rangle_{34}\right)/\sqrt{2}, \quad (27)$$

which is applied to the teleported qubit (10)

$$\Omega\left|\varphi\right\rangle_2 = \left|\Delta_1^{(id)}\right\rangle_{1234}. \quad (28)$$

Operator $\Omega$ is a superposition of the terms composed of the states and displacement operators with absolute equal but opposite in sign amplitudes. Further the state (21) is used in analysis of the quantum teleportation protocol of unknown qubit.

In general case of $t \neq 1$, the fidelities (25) are the functions of the amplitudes of the teleported state, its phase relationships, displacement amplitude $\alpha$ due to presence of the expansion coefficients $c_{nm}(\alpha)$ in expression for the fidelities as well as transmittance $t$. Numerical calculations show that the fidelities though depend on the parameters of the teleported state, this relationship is not significant. The fidelity is largely determined by two parameters $\alpha$ and $t$. Corresponding plot in figures 2 shows the fidelity (25) as function of $\alpha$ and $t$ for equal modulo amplitudes of the teleported qubit $a_0 = a_1 = 1/\sqrt{2}$. Similar dependencies little different from each other are observed for other amplitudes values. Realization of the displacement operator with help of HTBS is extremely sensitive to the parameters $\alpha$ and $t$. The fidelities take unit value in the case of $t=1$ regardless the displacement amplitude $\alpha$ [31]. But they fairly quickly fall almost to zero with increasing displacement parameter $\alpha$ and decreasing transmittance $t$. In order to achieve high fidelity $Fid(\alpha)>0.99$ it is required to choose a beam splitter with extremely high transmittance $t\cong 1$ or to consider displacement of the initial state on relatively small value $\alpha<1$. Thus, the efficiency of the displacement method with HTBS cannot be recognized to be high as the operation can only be effectively performed on low values of the displacement amplitudes $|\alpha|\leq 1$ for those values of transmittance $t$, which could be used in practice. We can hardly say that the generated entangled hybrid state (11) involve macroscopic (e. g., visible by eye [29,30]) states. Displaced entangled states of low amplitude were also used for the implementation of the dense coding protocol in [32].

## 3 Direct implementation of quantum teleportation protocol by superposition of displacement operators with absolute equal but opposite in sign amplitudes

In the previous section, we considered deterministic interaction of the quantum channel (11) with the teleported unknown qubit (10) on HTBS. As result of the interaction, the teleported qubit is displaced on values $\pm\alpha$ in such a way that all information on what quantity the teleported qubit has been displaced disappears. It is this uncertainty that underlies the successful implementation of the protocol. Ideal output state (21) can be rewritten in the terms of even/odd SCS as

$$BS\left(|\Psi\rangle_{134}|\varphi\rangle_2\right)=\sum_{n=0}^{\infty}g_n\begin{pmatrix}\frac{1}{N_+}|even\rangle_1\frac{N_n}{\sqrt{2}}\left((a_0+A_na_1)|01\rangle_{34}+(-1)^n(a_0-A_na_1)|10\rangle_{34}\right)+\\ \frac{1}{N_-}|odd\rangle_1\frac{N_n}{\sqrt{2}}\left((a_0+A_na_1)|01\rangle_{34}-(-1)^n(a_0-A_na_1)|10\rangle_{34}\right)\end{pmatrix}|n\rangle_2, \quad (29)$$

where the following parameters are introduced

$$A_n=\frac{n-\alpha^2}{\alpha}, \quad (30)$$

$$N_n=\left(|a_0|^2+A_n^2|a_1|^2\right)^{-0.5}=\left(1+\left(A_n^2-1\right)|a_1|^2\right)^{-0.5}, \quad (31)$$

$$g_n=F\frac{\alpha^n}{2N_n\sqrt{n!}}. \quad (32)$$

$$|even\rangle_1=N_+\left(|0,-\beta\rangle+|0,\beta\rangle\right)_1, \quad (33a)$$

$$|odd\rangle_1=N_-\left(|0,-\beta\rangle-|0,\beta\rangle\right)_1, \quad (33b)$$

where the factors $N_\pm=\left(2\left(1\pm\exp\left(-2|\beta|^2\right)\right)\right)^{-1/2}$ are the normalization parameters of the even/odd SCS. Here, amplitude $\alpha$ is real quantity (22).

After that, Alice performs two types of measurements: parity measurement in first mode and photon number measurement in the second mode. Having performed parity measurement (even/odd) in the first mode and photon number resolving detection in the second mode, Alice generates the following two states at Bob's disposal

$$\frac{N_n}{\sqrt{2}}\left((a_0+A_na_1)|01\rangle_{34}+(a_0-A_na_1)|10\rangle_{34}\right),\text{if } (j=0,n=2m) \text{ and } (j=1,n=2m+1), \quad (34a)$$

$$\frac{N_n}{\sqrt{2}}\left((a_0+A_na_1)|01\rangle_{34}-(a_0-A_na_1)|10\rangle_{34}\right),\text{if } (j=0,n=2m+1) \text{ and } (j=1,n=2m), \quad (34b)$$

where the notation $(j=0/j=1,n=2m/2m+1)$ indicates that *even/odd* number of photons is registered in the first mode and $n=2m$ or $n=2m+1$ photons is fixed in the second mode by Alice. Value $j=0$ corresponds to even SCS and $j=1$ encodes odd SCS. For example, record $(j=0,n=2m)$ in (34a) implies that Alice has detected even SCS by measuring the *coherent* mode and $n$ photons in the second mode where $n=2m$ is even number. Alice can record the results of her measurements by a string of two digits $(j,n)$. The first number $j$ is the binary one and displays the result of her parity measurement, while the second number $n$ is a decimal and displays the number of measured photons in the teleported qubit. A decimal number $n$ must be translated into binary code before sending the message to Bob increasing total bit line length. Bob reads message and realizes the Pauli $Z-$operation $Z^{j+par(n)}$ on his dual-rail single photon by phase shift in one of two modes by $\pi$, where $j=0,1$ and $par(n)$ means the parity of number $n$. After this, Bob performs Hadamard operation $H$ on his dual-rail single photon regardless of Alice's measurement outcomes

$$HZ^{j+par(n)}\frac{N_n}{\sqrt{2}}\begin{vmatrix} a_0+A_na_1 \\ (-1)^j(a_0-A_na_1) \end{vmatrix}=|\psi_n\rangle=N_n\begin{vmatrix} a_0 \\ A_na_1 \end{vmatrix}, \tag{35}$$

where $H$ is Hadamard transformation

$$H=\frac{1}{\sqrt{2}}\begin{vmatrix} 1 & 1 \\ 1 & -1 \end{vmatrix}, \tag{36a}$$

and $Z$ matrix is

$$Z=\begin{vmatrix} 1 & 0 \\ 0 & -1 \end{vmatrix}. \tag{36b}$$

The Hadamard operation on a single photon can be implemented using the balanced beam splitter and phase shift operations [33]. The states $|\psi_n\rangle$ (Eq. (35)) is written in dual-rail basis of single photon $\{|01\rangle,|10\rangle\}$ unlike initial $\{|0\rangle,|1\rangle\}$. The states are not original (10) as they involve additional factors $N_n$ and $N_nA_n$, respectively. But the states $|\psi_n\rangle$ conserve phase relations with original one. Since the states include additional factors known to Alice and not affecting the phase relations, then such states can be called amplitude-modulated (AM). Subscript $n$ corresponds to number of photons measured in the teleported state and defines parameter of amplitude modulation (30) of the output state. The reverse process can be called the demodulation of the output stats and is not considered in the present work.

The success probability to generate the states (35) at Bob's station depends on parameters of the teleported state, namely, on the absolute values of the amplitudes $|a_0|$ or $|a_1|$

$$P_n(\alpha)=\exp\left(-|\alpha|^2\right)\frac{|\alpha|^{2n}\left(1+|a_1|^2\left(A_n^2-1\right)\right)}{n!}. \tag{37}$$

One can directly show using (A20) that the total probability of the events is equal to one $\sum_{n=0}^{\infty}P_n(\alpha)=1$. Note that the additional amplitude factors and, as consequence, the dependence of the success probability on the absolute values of the teleported qubit arise as a result of the fact that the coherent state and displaced single photon state are transformed differently when projecting the state onto measurement basis of the number states. Corresponding three-dimensional plots $P_n$ for $n=0,1,2,3$ as functions of $\alpha$ and $|a_1|$ are presented in Fig. 3(a-d). Variation range of the displacement amplitude is chosen within $\alpha\in[-0.5,0.5]$. Plots in figure 3(a-d) show that the probabilities $P_0$ and $P_1$ prevail over other probabilities (especially over the probabilities $P_n$ of higher order with $n>3$) in a wide range of change of the displacement amplitude $\alpha$ and absolute values of the qubit amplitude $|a_1|$. These plots allow us to claim the less we choose the value of the displacement amplitude, the greater we observe the preponderance of $P_0$ and $P_1$ over the other probabilities $P_n$ with $n>1$ $(P_0,P_1>>P_n)$. The plot in figure 3(e) made for $\alpha=0.03$ fully supports this conclusion. In the case, the sum probability takes a minimum value $P_0+P_1\geq P_{\min}=0.9982$. It is worth noting that the probabilities $P_0$ and $P_1$ evolve in opposite relation to each other, that is, if $P_0$ falls, then $P_1$ increases, and vice versa. The probabilities of higher orders $P_n$ with $n>1$ take much smaller values in the entire range of variation of the absolute amplitude $|a_1|$.

It is unlikely that such implementation of the quantum teleportation protocol of unknown qubit can be practical since it requires the use of special detectors capable to determine the parity of even/odd SCS and discriminate among incoming photons. In addition, an increase in the measurement outcomes leads to an increase of the classical information flow from Alice to Bob which can be hardly considered an advantage of the approach. Therefore, reducing the displacement amplitude $\alpha$ becomes the best strategy in terms of implementation of the studied protocol in practice. Moreover, the amplitude of the coherent components of the hybrid channel (11) can be significantly reduced. For example, consider the beam splitter (1) with transmission coefficient $T=t^2=0.99$ which in combination with the small displacement amplitude value $\alpha=0.03$ gives small value $\beta=0.3$ (Eq. (22)) of amplitude of the coherent components. Use of the SCS with the small amplitude value allows us to approximate them by the following number states with high fidelity

$$|even\rangle \approx |0\rangle, \tag{38a}$$

$$|odd\rangle \approx |1\rangle. \tag{38b}$$

Indeed, the probability distributions of the *even/odd* SCS, for example, with $\beta=0.3$

$$P_{2n}^{even}(\beta)=4N_+^2(\beta)\exp\left(-|\beta|^2\right)\left(|\beta|^2\right)^{2n}/(2n)!, \tag{39a}$$

$$P_{2n+1}^{odd}(\beta)=4N_-^2(\beta)\exp\left(-|\beta|^2\right)\left(|\beta|^2\right)^{2n+1}/(2n+1)!, \tag{39b}$$

take the following values

$P_0^{even}(\beta=0.3)=0.996$, $P_2^{even}(\beta=0.3)=0.004$, $P_4^{even}(\beta=0.3)=2.72\times10^{-6}$,
$P_1^{odd}(\beta=0.3)=0.9986$, $P_3^{odd}(\beta=0.3)=0.0013$, $P_5^{even}(\beta=0.3)=4.9\times10^{-8}$.

Probabilities $P_0^{even}(\beta=0.3)$ and $P_0^{odd}(\beta=0.3)$ prevail over other ones that enables to make use of approximation (38). This is more than enough to take advantage of commercially achievable avalanche photodiode (APD) being a highly sensitive semiconductor electronic device that exploits the photoelectric effect to convert light to electricity and that can ideally operate in on-off regime $|0\rangle\langle 0|+\sum_{n=1}^{\infty}|n\rangle\langle n|$. Thus, the parity measurement in the case of $\beta<1$ can be replaced by APD able to distinguish outcomes from vacuum and single photon. Use of the small amplitudes $\beta<1$ of the quantum channel (11) guaranties performance of $\alpha<<1$ (22). Then, registration of two photons (not mentioning light pulses with a larger number of photons) in the teleported mode is unlikely as the probability of such events is less than one percent (Figs. 3(a-d)). This means that APD can also be used in the teleported mode to get measurement outcomes instead of photon number resolving detector in the case of $\alpha<<1$.

Finally, the quantum teleportation protocol of unknown qubit can be described in simpler form in the case of $\alpha<<1$ instead of (29)

$$BS\left(|\Psi\rangle_{134}|\varphi\rangle_2\right)=g_0\begin{pmatrix}\frac{1}{N_+}|00\rangle_{12}\frac{N_0}{\sqrt{2}}\left((a_0+A_0a_1)|01\rangle_{34}+(a_0-A_0a_1)|10\rangle_{34}\right)+\\ \frac{1}{N_-}|10\rangle_{12}\frac{N_0}{\sqrt{2}}\left((a_0+A_0a_1)|01\rangle_{34}-(a_0-A_0a_1)|10\rangle_{34}\right)\end{pmatrix}+$$
$$g_1\begin{pmatrix}\frac{1}{N_+}|01\rangle_{12}\frac{N_1}{\sqrt{2}}\left((a_0+A_1a_1)|01\rangle_{34}-(a_0-A_1a_1)|10\rangle_{34}\right)+\\ \frac{1}{N_-}|11\rangle_{12}\frac{N_1}{\sqrt{2}}\left((a_0+A_1a_1)|01\rangle_{34}+(a_0-A_1a_1)|10\rangle_{34}\right)\end{pmatrix}, \tag{40}$$

with fidelity prevailing $>0.99$. Subsequent Alice's measurement in the base $\{|0\rangle,|1\rangle\}$ instead of $\{|even/odd\rangle,|n\rangle\}$ allows Bob to receive one of two possible states either $|\psi_0\rangle$ or $|\psi_1\rangle$ (Eq. (35)). Note only that Alice's measured outcomes are the same as if she had performed Bell-state measurement $|00\rangle_{12}, |10\rangle_{12}, |01\rangle_{12}$, and $|11\rangle_{12}$, respectively [7,24]. Bob does not know exactly which state he has at his disposal and he needs Alice help to identify them. Alice encodes your data two bits in length as in [7] in full compliance with her measured results $|jk\rangle_{12} \rightarrow (j,k)$, where $j,k=0,1$ (Fig. 1). Bob reads the bits and decides whether or not to apply the $Z$ operation $(Z^{j+k})$ to his single photon with subsequent application of the Hadamard operation. The second bit is assigned for Bob to unambiguously determine which state either $|\psi_0\rangle$ $(k=0)$ or $|\psi_1\rangle$ $(k=1)$ he obtained. Summarizing all of the above, the results of the chapter are reflected in Table 1 for the case of $\alpha<<1$.

| Measurement outcomes | Teleported state | Success probability |
|---|---|---|
| $(0,0)$, $(1,0)$ | $\|\psi_0\rangle$ (AM) | $P_0(\alpha)$ |
| $(0,1)$, $(1,1)$ | $\|\psi_1\rangle$ (AM) | $P_1(\alpha)$ |

**Table 1**. Concise generalization of quantum teleportation protocol with hybrid channel (11). AM means amplitude-modulated state.

### 4 Increase of efficiency of the protocol by input amplitude modulation of unknown qubit

We have shown the use of small values of the displacement amplitude $\alpha$ on which we need to displace the teleported qubit allows us significantly to increase effectiveness of the protocol. Parity measurement and photon number resolving measurement can be replaced by on-off measurement that greatly increases the chances to implement the protocol in practice. Hybrid quantum channel (11) with a sufficiently small value of the amplitude of the coherent states can be used in practice which is also an advantage of the protocol. Nevertheless, the problem of amplitude demodulation of the output qubits remains. To increase the efficiency of the protocol let us embrace initial amplitude modulation of the original qubit (10) as

$$\left|\varphi_0^{(in)}\right\rangle_2 = N_0^{(in)} \begin{vmatrix} a_0 \\ A_0^{-1} a_1 \end{vmatrix}, \tag{41}$$

where amplitudes $a_0$ and $a_1$ are unknown parameters, $A_0$ is a modulation factor given by (30) and the normalization factor $N_0^{(in)}$ is given by $N_0^{(in)} = \left(|a_0|^2 + |A_0|^{-2}|a_1|^2\right)^{-0.5} = \left(1+\left(|A_0|^{-2}-1\right)|a_1|^2\right)^{-0.5}$. The protocol is also performed as described in Section 2 and the results are shown in Table 2 in the case of $\alpha<<1$.

| Measurement outcomes | Teleported state | Success probability |
|---|---|---|
| $(0,0)$, $(1,0)$ | $\|\psi_{00}^{(out)}\rangle$ (original) | $P_{00}(\alpha)$ |
| $(0,1)$, $(1,1)$ | $\|\psi_{10}^{(out)}\rangle$ (AM) | $P_{10}(\alpha)$ |

**Table 2**. Results of quantum teleportation protocol of initial AM unknown qubit (41) in the case of $\alpha<<1$.

Here, the following notations are introduced

$$\left|\psi_{00}^{(out)}\right\rangle = \begin{vmatrix} a_0 \\ a_1 \end{vmatrix}, \tag{42}$$

$$\left|\psi_{n0}^{(out)}\right\rangle = N_{n0}\begin{vmatrix} a_0 \\ A_n A_0^{-1} a_1 \end{vmatrix}, \tag{43}$$

where the normalization factor is $N_{n0} = \left(|a_0|^2 + |A_n|^2|A_0|^{-2}|a_1|^2\right)^{-0.5} = \left(1 + \left(|A_n|^2|A_0|^{-2} - 1\right)|a_1|^2\right)^{-0.5}$. Corresponding success probabilities are given by

$$P_{00}(\alpha) = \exp\left(-|\alpha|^2\right)N_0^{(in)2} = \frac{\exp\left(-|\alpha|^2\right)}{\left(1 + \left(|A_0|^{-2} - 1\right)|a_1|^2\right)}, \tag{44}$$

$$P_{n0}(\alpha) = \exp\left(-|\alpha|^2\right)\frac{|\alpha|^{2n}}{n!}\frac{N_0^{(in)2}}{N_{n0}^2} = \exp\left(-|\alpha|^2\right)\frac{|\alpha|^{2n}}{n!}\frac{1 + \left(|A_n|^2|A_0|^{-2} - 1\right)|a_1|^2}{1 + \left(|A_0|^{-2} - 1\right)|a_1|^2}, \tag{45}$$

where $n = 1$ is taken in Table 2.

The Alice's choice of amplitude modulation of the teleported qubit (41) allows to Bob to immediately get the initial state (42) with probability (44). The distribution (44,45) differs one (37). It is possible directly to show the distribution (44,45) is normalized $\sum_{n=0}^{\infty} P_{n0}(\alpha) = 1$. Plots of the probability dependencies $P_{00}$, $P_{10}$ and $P_{20}$ on $|a_1|$ are shown in figures 4(a-d) for different values of the displacement parameter $\alpha$. As can be seen from the plots, there is a range of values of $|a_1|$, at which the success probability of teleportation (41) is more of $0.5$. Participants of the teleportation protocol (Alice and Bob) may be fortunate even not suspecting about it and teleport unknown qubit (10) with success probability close to one if highly unbalanced qubit with $|a_0| << |a_1|$ is used. Increase of the displacement parameter $\alpha$ enables to increase the range of values $|a_1|$ for which the success probability of the teleportation is more than 0.5 (Figs. 4(c,d)). But the increase of the displacement amplitude $\alpha$ is restricted from a practical point of view. Contribution of the state $\left|\psi_{20}^{(out)}\right\rangle$ (curve 3 in Fig. 4(c,d)) may become essential in the case of increase of the displacement amplitude $\alpha$.

Consider another type of amplitude modulation of the unknown qubit (10)

$$\left|\varphi_1^{(in)}\right\rangle_2 = N_1^{(in)}\begin{vmatrix} a_0 \\ A_1^{-1} a_1 \end{vmatrix}, \tag{46}$$

where quantity $A_1$ is the modulation factor given by (30) and the normalization factor $N_1^{(in)}$ is $N_1^{(in)} = \left(|a_0|^2 + |A_1|^{-2}|a_1|^2\right)^{-0.5} = \left(1 + \left(|A_1|^{-2} - 1\right)|a_1|^2\right)^{-0.5}$. Results of the protocol are presented in Table 3.

| Measurement outcomes | Teleported state | Success probability |
|---|---|---|
| $(0,0)$, $(1,0)$ | $\left\|\psi_{01}^{(out)}\right\rangle$ (AM) | $P_{01}(\alpha)$ |
| $(0,1)$, $(1,1)$ | $\left\|\psi_{11}^{(out)}\right\rangle$ (original) | $P_{11}(\alpha)$ |

**Table 3**. Results of quantum teleportation protocol of initial AM unknown qubit (46) in the case of $\alpha << 1$.

Here, we use the following notations

$$\left|\psi_{11}^{(out)}\right\rangle = \begin{vmatrix} a_0 \\ a_1 \end{vmatrix}, \tag{47}$$

$$\left|\psi_{n1}^{(out)}\right\rangle = N_{n1}\begin{vmatrix} a_0 \\ A_n A_1^{-1} a_1 \end{vmatrix}, \tag{48}$$

where the normalization factor is $N_{n1} = \left(|a_0|^2 + |A_n|^2 |A_1|^{-2} |a_1|^2\right)^{-0.5} = \left(1 + \left(|A_n|^2 |A_1|^{-2} - 1\right)|a_1|^2\right)^{-0.5}$ and $n \neq 1$. The success probabilities are the following

$$P_{11}(\alpha) = \exp\left(-|\alpha|^2\right)|\alpha|^2 N_1^{(in)2} = \frac{\exp\left(-|\alpha|^2\right)|\alpha|^2}{\left(1 + \left(|A_1|^{-2} - 1\right)|a_1|^2\right)}, \tag{49}$$

$$P_{n1}(\alpha) = \exp\left(-|\alpha|^2\right)\frac{|\alpha|^{2n}}{n!}\frac{N_1^{(in)2}}{N_{n1}^2} = \exp\left(-|\alpha|^2\right)\frac{|\alpha|^{2n}}{n!}\frac{1 + \left(|A_n|^2 |A_1|^{-2} - 1\right)|a_1|^2}{1 + \left(|A_1|^{-2} - 1\right)|a_1|^2}, \tag{50}$$

where $n = 0$ is taken in Table 3. This third distribution is normalized $\sum_{n=0}^{\infty} P_{n1}(\alpha) = 1$ and different from two others (37, 44,45). Plots of the probabilities $P_{01}$, $P_{11}$ and $P_{21}$ in dependency on $|a_1|$ are shown in figures 5(a-d) for different values of the displacement parameter $\alpha$. As well as for the distribution (44,45), there is the range of amplitude values $|a_1|$ for which the teleportation of unknown qubit (10) occurs with a success probability greater than $0.5$. This range is shifted to higher amplitude values $|a_1| \cong 1$ in contrast to the case of amplitude modulation (41).

So, Alice can use the technique of initial amplitude modulation of the unknown qubit with aim to teleport to Bob two states one of which is input (10) in the basis of $\{|01\rangle, |10\rangle\}$. Technique of initial amplitude modulation of an unknown qubit allows us to implement quantum teleportation protocol with probability of success bigger than 0.5 but only for two cases, when the teleported qubit is significantly unbalanced either with $|a_1| < 0.4$ (Figs. 4) or $|a_1| > 0.95$ (Figs. 5). Here, the problem of amplitude demodulation becomes relevant for one of the states (either $\left|\psi_{10}^{(out)}\right\rangle$ or $\left|\psi_{01}^{(out)}\right\rangle$) in order to achieve nearly deterministic implementation of the protocol of quantum teleportation of unknown qubit unlike the case of direct teleportation (Table 1). Therefore, this strategy with the initial amplitude modulation can look more preferable compared with the case discussed in the previous section as it guarantees exact teleportation with some probability and decreases number of the states requiring amplitude demodulation (one state instead of two). Moreover, if the teleported qubit is highly unbalanced, then the success probability of the protocol may become close to one provided that Alice successfully guessed with amplitude modulation of the qubit. Note also that Bob unambiguously knows which of two states he obtained either original (42,47) or AM (43,48) after Alice has sent him auxiliary classical information in length of $2$ bits.

Consider another strategy aimed at increasing the efficiency of the protocol. Alice can prepares AM qubit either

$$\left|\varphi^{(0)}\right\rangle_2 = \begin{vmatrix} a_0^{(0)} \\ a_1^{(0)} \end{vmatrix} = N^{(0)}\begin{vmatrix} a_0 \\ b_0 a_1 \end{vmatrix}, \tag{51a}$$

or

$$\left|\varphi^{(1)}\right\rangle_2 = \begin{vmatrix} a_0^{(1)} \\ a_1^{(1)} \end{vmatrix} = N^{(1)}\begin{vmatrix} a_0 \\ b_1 a_1 \end{vmatrix}, \tag{51b}$$

where the normalization condition $\left|a_0^{(0)}\right|^2 + \left|a_1^{(0)}\right|^2 = \left|a_0^{(1)}\right|^2 + \left|a_1^{(1)}\right|^2 = 1$ is performed for both qubits and $N^{(0)} = \left(|a_0|^2 + |b_0|^2 |a_1|^2\right)^{-1/2}$, $N^{(1)} = \left(|a_0|^2 + |b_1|^2 |a_1|^2\right)^{-1/2}$ are the normalization factors. The

factors $b_0 <<1$ and $b_1 >>1$ are chosen in such a way to ensure that the teleported qubit is highly unbalanced. Now, she decides to teleport such unbalanced qubit instead of (10). Despite the fact that Alice has partial access to information about teleported qubits (the meanings of $b_0$ and $b_1$ which she assigns herself) in general the qubits (51a) and (51b) remain completely unknown ($a_0$ and $a_1$ are unknown values). Now, using data of Tables 2 and 3 and plots in Figs. 4 and 5, Alice can purposefully once more modulate her unknown qubit as either

$$\left|\varphi_0^{(0)}\right\rangle_2 = N_0^{(in)}\left|\begin{matrix} a_0^{(0)} \\ A_0^{-1}a_1^{(0)} \end{matrix}\right|, \tag{52a}$$

or

$$\left|\varphi_1^{(1)}\right\rangle_2 = N_1^{(in)}\left|\begin{matrix} a_0^{(1)} \\ A_1^{-1}a_1^{(1)} \end{matrix}\right|. \tag{52b}$$

Now, Alice teleports amplitude modulated unknown qubit and Bob obtains at his disposal one of the states (51) in base of $\{|01\rangle, |10\rangle\}$. If Alice was right in her hypothesis concerning the qubits (52a) and (52b), then Bob may almost deterministically (or at least with success probability more of 0.5) get one of the teleported states either (52a) or (52b) in dependency on Alice's conjecture after finishing the quantum teleportation protocol in Fig. 1 in regime of $\alpha <<1$ as described above. Results presented in Table 2 and 3 as well as plots in Figs. 4,5 are applicable to the case provided that replacements either $a_0 \to a_0^{(0)}, a_1 \to a_1^{(0)}$ or $a_0 \to a_0^{(1)}, a_1 \to a_1^{(0)}$ are used. Alice has full information about how much the amplitudes of the teleported qubit are changed and she can even share it with Bob by a separate message. The states (51a) and (51b) are not the same as initially unknown qubit (10). Nevertheless, the states have the same phase relations as original (10). The initial amplitude modulation of unknown qubit (51) (either $b_0 <<1$ or $b_1 >>1$) may not be sufficient to satisfy the condition of highly unbalanced qubit and guarantee a high probability of the protocol. Here, we can only talk about blind amplitude modulation of the unknown qubit in the case. To evade it Alice can perform the protocol in another interpretation. Suppose Alice has access to the amplitude information $|a_0|$, $|a_1|$ not knowing anything about the phase relations of the qubit (10). Then, she can choose the relevant factors to ensure generation of unbalanced AM unknown qubit either (51a) or (51b). We can talk about deliberate amplitude modulation of unknown qubit in the case. Note that Alice can make amplitude modulation in one step simultaneously instead of doing it twice. Alice can easy calculate the additional factor either $c_0 = b_0 A_0^{-1}$ or $c_1 = b_1 A_1^{-1}$ and do amplitude modulation of her unknown qubit (10) in one step. Note Bob can embrace the received AM state either $|\psi_0\rangle$ or $|\psi_1\rangle$ to teleport it to a third party to Charles in the same manner as shown in Fig. 1 by means of preliminary amplitude modulation like (52a) and (52b). In the case, Bob's highly unbalanced state either (51a) or (51b) is original which can be nearly deterministically teleported to Charles.

If Alice has access to the amplitude information about the state not knowing anything about the phase relations of the qubit, she can choose the relevant factors (either $b_0 <<1$ or $b_1 >>1$) in order to ensure the greatest possible success probability of the quantum teleportation of amplitude modulated unknown qubit (51). Note that even partial knowledge about qubit (information about amplitudes) leaves the qubit unknown. Thus, Alice and Bob have a possibility to implement quantum teleportation protocol of partially known qubit in nearly deterministic manner provided that Alice initially modulates it as given by (51). Note the case in not remote state preparation (RSP) [35-38]. RSP is a quantum communication protocol which can be considered as a kind of quantum teleportation that allows indirect transfer of quantum information between two distant parties by means of a shared entangled resource and classical information. In the protocol, Alice does not possess a copy of the source state, but she is aware of its full classical description.

## 5 Generation of the hybrid quantum channel

The same mechanism can be used to generate the hybrid channel (11). Consider interaction of even SCS (33a) with maximally entangled state of two photons in superposition of two modes

$$|\varphi\rangle_{3456} = |0101\rangle_{3456} + |1010\rangle_{3456}, \tag{53}$$

as shown in Fig. 6. The even SCS state interacts with fifth mode of the state (53) on HTBS (1). Additionally, the coherent state $|-\beta_1\rangle_2$ occupying second mode interacts with its mode 6 on another HTBS

$$\begin{aligned} &BS_{15}BS_{26}\left(|even\rangle_1|-\beta_1\rangle_2|\varphi\rangle_{3456}\right) = \\ &N_+\left(a_0BS_{15}BS_{26}\left(|-\beta\rangle_1|-\beta_1\rangle_2|\varphi\rangle_{3456}\right) + a_1BS_{15}BS_{26}\left(|\beta\rangle_1|-\beta_1\rangle_2|\varphi\rangle_{3456}\right)\right)\end{aligned}. \tag{54}$$

Following the mathematical approach developed in section 2, we can obtain

$$BS_{15}BS_{26}\left(|even\rangle_1|\beta_1\rangle_2|\varphi\rangle_{3456}\right) \to \left|\Delta_1^{(id)}\right\rangle_{13456}|0,-\beta_1/t\rangle_2, \tag{55}$$

in the limit case of $t=1,\ r=0$, where the ideal normalized state is the following

$$\left|\Delta_1^{(id)}\right\rangle_{13456} = N_+F^2\sum_{n=0}^{\infty}\sum_{m=0}^{\infty}|\Psi_{nm}\rangle_{134}|nm\rangle_{56}, \tag{56}$$

where the following states are introduced

$$|\Psi_{nm}\rangle_{134} = |0,-\beta\rangle_1\left|\varphi_{nm}^{(+)}(\alpha,\alpha_1)\right\rangle_{34} + (-1)^n|0,\beta\rangle_1\left|\varphi_{nm}^{(-)}(\alpha,\alpha_1)\right\rangle_{34}, \tag{57}$$

with

$$\left|\varphi_{nm}^{(\pm)}(\alpha,\alpha_1)\right\rangle_{34} = \left(c_{0n}(\alpha)c_{1m}(\alpha_1)|01\rangle_{34} \pm c_{1n}(\alpha)c_{0m}(\alpha_1)|10\rangle_{34}\right)/\sqrt{2}. \tag{58}$$

Deterministic displacement of the state (53) by the values $\pm\alpha$ by the coherent components of even SCS with disappearing all information about the events is followed by a probabilistic measurement in auxiliary modes 5 and 6 in Fig. 6. Let us consider the case $\alpha=\alpha_1$. We are interested in registration of events either $|01\rangle_{56}$ or $|01\rangle_{56}$ that gives birth to the states

$$|\Psi_{01}\rangle_{134} = \left(|0,-\beta\rangle_1\left(|01\rangle_{34}+|10\rangle_{34}\right) + |0,-\beta\rangle_1\left(|01\rangle_{34}-|10\rangle_{34}\right)\right)/\sqrt{2}, \tag{59}$$

and

$$|\Psi_{10}\rangle_{134} = \left(|0,-\beta\rangle_1\left(|01\rangle_{34}+|10\rangle_{34}\right) - |0,-\beta\rangle_1\left(|01\rangle_{34}-|10\rangle_{34}\right)\right)/\sqrt{2}. \tag{60}$$

Application of Hadamard gate (36a) with subsequent $Z^n$ (36b), where $n=0,1$, generates the input quantum channel (11). It can be shown that, as the case of the quantum teleportation protocol, the probability of the measurement outcomes $(00)$, $(01)$, $(10)$ and $(11)$ significantly prevails over higher order measurement events $(mn)$ in the case of $\alpha<<1$ which makes possible to use two APD (Fig. 6).

Let us consider the possibility of realization of amplitude modulation [39] of an unknown qubit (10). Note that amplitude modulation is preliminary and preparatory operation not directly included in the protocol of quantum teleportation. It is assumed that the AM unknown qubit (41,46) is initially ready before the teleportation. Nevertheless, the same mechanism associated with displacement of unknown qubit can be used and even be involved into quantum teleportation protocol. Consider interaction of unknown qubit (dual-rail single photon)

$$|\varphi_{In}\rangle_{23} = a_0|01\rangle_{23} + a_1|10\rangle_{23}. \tag{61}$$

with coherent state $|0,-\beta\rangle_1$ on HTBS (1) as shown in Figs. 1,6. Then, we have the following chain of transformations

$$\begin{aligned} &BS_{13}\left(|0,-\beta\rangle_1|\varphi_{In}\rangle_{23}\right) \to |0,-\beta\rangle_1\left(a_0|0\rangle_2|1,\alpha\rangle_3 + a_1|1\rangle_2|0,\alpha\rangle_3\right) = \\ &F|0,-\beta\rangle_1\sum_{n=0}^{\infty}\left(a_0c_{1n}(\alpha)|0\rangle_2 + a_1c_{0n}(\alpha)|1\rangle_2\right)|n\rangle_3\end{aligned}. \tag{62}$$

Subsequent measurement of $n$ photons in the third mode projects the state (62) onto one of AM states. Consider the case of $\alpha<<1$. Then, success probability to register vacuum and single photon in third mode prevails over other measurement outcomes. Indeed, the success probabilities are the following

$$P_{In0} = F^2 |\alpha|^2 N_0^{(in)^2}, \tag{63a}$$

$$P_{In1} = F^2 \left(1 - |\alpha|^2\right)^2 N_1^{(in)^2}, \tag{63b}$$

where the parameters $N_0^{(in)}$ and $N_1^{(in)}$ are given above. The sum of the probabilities can be approximated by

$$P_{In0} + P_{In1} \approx F^2 \left(1 + |\alpha|^2 \left(|a_1|^2 - |a_0|^2\right)\right) \to 1, \tag{64}$$

when $\alpha \to 0$. Note only that Alice obtains AM state (41) ready for the teleportation when she registers vacuum. If Alice measures a single photon, then she obtains another AM state (46) prepared for the teleportation. Thus, protocol of quantum teleportation of AM state (either (41) or (46)) can be implemented nearly deterministically even taking into account the preparation procedure of the teleported qubit.

## 6 Results

We considered mechanism of interaction of continuous variable states with discrete variable states to increase efficiency of the quantum teleportation protocol of unknown qubit. It is considered the interaction on HTBS with aim to realize action of the displacement operator on target state [31]. Here, we developed novel way to implement quantum teleportation protocol of unknown qubit without use of Bell state formalism [24]. Coherent components of the hybrid channel (11) simultaneously displace unknown teleported qubit in indistinguishable manner on HTBS by the values that differ from each other only by sign $\pm\alpha$. Both coherent components of the quantum channel displace the teleported state so that we do not have access to information on what value (either $\alpha > 0$ or $\alpha < 0$) the teleported qubit is displaced despite the fact that one of these displacements has already happened. This is uncertainty is key moment for implementation of the quantum teleportation protocol. Projection of the uncertain teleported state being superposition of components from different Hilbert spaces determined by parameters $\alpha$ and $-\alpha$ onto measurement basis produces desired controlled$-Z$ operation due to relation (A13) and teleported state can be recovered. We have shown the interaction between the quantum channel (11) and unknown teleported qubit (10) on HTBS can be approximated by the operator composed of the displacement operators with absolute equal but opposite in sign amplitudes (27) acting on the target state (28). Mathematical apparatus is developed and confirms the correctness of the assertion. Numerical calculations of fidelity presented in Fig. 2 show the deterministic indistinguishable displacement of the teleported qubit is especially effectively implemented in the case of $\alpha << 1$. Implementation of the displacement operators for large values of $\alpha$ looks vague and requires very extreme values of transmittance $t$. Note only the same mechanism can be applied for generation of the quantum channel and amplitude modulation of unknown qubit.

We have shown the implementation can be especially effectively performed in the case of small values $\pm\alpha$ on which the teleported qubit can be displaced with 2 bits of classical information to identify all states (Table 1). Measurement of the number of photons and the parity of the SCS states can be replaced by on-off measurement (there is or no photon in mode) that significantly reduces the technical difficulties in the practical implementation of the protocol. Two commercially achievable APDs are enough for measurements in practical implementation of the quantum teleportation protocol in the case of $\alpha << 1$. This case has serious advantages in addition to reducing the number of measured photons. The amplitude (Eq. (11)) of the quantum channel and, as consequence of SCS, is significantly reduced. SCS of such size can be produced in practice [25-27,40]. Increase of the displacement amplitude $\alpha$ entails both increase of the received information and use of a superconducting single-photon detector [41] (SSPD) working at cryogenic temperature. The transformations ($Z$ and $H$ transformations) over a single photon in superposition state of two modes can be efficiently produced by the linear optical elements. In general, the protocol is realized by linear optics methods and APDs.

Despite its simplicity and availability, there is a price that has to be paid. The resulting qubit in Bob's hands becomes amplitude modulated but conserves the same phase relation with original teleported one. Amplitude modulation or amplitude distortion of the output state on compared with original arises due to displaced vacuum (coherent state) and displaced single photon are transformed differently from each other under projection onto measurement basis. Amplitude modulation of the output states is an inalienable and inherent feature of the protocol as an inability to implement complete Bell-state measurement by linear optics methods [8]. But this obstacle is bypassed by the choice of the appropriate strategy by Alice. Instead of the direct implementation of the quantum teleportation protocol, Alice can make use of the method of initial amplitude modulation to increase the effectiveness of the protocol. Note that, at least, Alice has complete information about the amplitude factors. Use of this method allows Alice to immediately restore the original qubit (Tables 2 and 3). The states that need to demodulate is decreased twice (one instead of two) in the considered case of $\alpha << 1$. Suppose that the teleported qubit was highly unbalanced and Alice guessed with correct amplitude modulation. Then, Bob obtains original (not AM) unknown qubit at his disposal with a probability of success close to unity or at least more than 50 percent. Assume she declares its AM unknown qubit as original (51). Then, Alice must once again modulate its amplitude (52) in order to prepare it for the teleportation. In reality, the amplitude modulations occur in one action and the factor that determines the amplitude modulation is divided in a corresponding proportion into two parts. If the condition of high unbalance for the AM unknown qubit is performed, then Bob obtains the desired state (which Alice already defined as the initial state) with a probability of success close to unity or at least more than 50 percent. It happens in the overwhelming majority of cases. To avoid possible failures, one can assume the case that Alice has a partial access to information about the teleported qubit, namely its amplitudes, without knowing anything about its phase relationships. Then, she can make meaningful amplitude modulation of the unknown qubit and teleport AM qubit (which is already recognized as the original one) in nearly deterministic manner. We cannot even consider that preparing AM qubit is auxiliary function that is not directly included in the procedure of the quantum teleportation. Since it is proved that amplitude modulation can be realized with a probability of success close to one, the procedure can be also included in the protocol.

Studied protocol of the quantum teleportation could be deterministic provided that we can find a way to definitely convert the states (35) to original (10) not having access to any information about qubit only knowing the amplitude factors. It is known the conversion could be implemented probabilistically. There are several promising ways to demodulate the unknown qubit that are currently under study. Solution of the problem of deterministic demodulation of unknown qubit remains open and it is transferred beyond the studied mechanism of the teleportation. The developed protocol of quantum teleportation performed by superposition of the displacement operators with opposite in phase amplitudes is different from previously proposed [7,16,20,42,43]. So, the protocol is free from those difficulties that that arise in the implementation of Bell state measurements by methods of linear optics [8]. Limit in 50 percent of success probability imposed by linear optics is not relevant to the proposed scheme. In proposed implementation, Alice can distinguish all her measurement outcomes in the case of $\alpha << 1$ and Bob knows exactly what state he has in his hands. Complete Bell state measurement can be done for the superpositions of entangled coherent states (SECS) [16]. But implementation of such a protocol is hardly possible in practice, since it requires SECS of large amplitude and PNRD. Use of hybrid entanglement can improve the efficiency of the protocol of quantum teleportation due to the possibility for Bob to realize $X$ operation [20]. But the same difficulties remain for Alice. Nearly deterministic realization of quantum teleportation protocol is possible through increase of used resources [42,43]. But such methods can hardly be used in practice due to the complexity in realization of the used quantum channels and significant increase of the optical elements required for implementation of measurement. Unlike them, the studied protocol can be implemented with a hybrid coherent entangled state (11) whose coherent components have small amplitudes. The studied protocol does not require auxiliary photons, complex quantum channels and hyperentanglement and can be implemented with irreducible number of optical elements: one beam splitter and two APDs (Fig. 1). Realization of quantum channel is performed in the same manner and requires additional beam splitter (Fig. 6).

## Appendix A. Properties of the displacement operator under change of its amplitude on opposite in sign

Unitary displacement operator [34] is determined by

$$D(\alpha)=\exp\left(\alpha a^{+}-\alpha^{*}a\right), \tag{A1}$$

where $\alpha$ is an amplitude of the displacement and $a$, $a^{+}$ are the bosonic annihilation and creation operators. Its action on number state results in

$$|n,\alpha\rangle = D(\alpha)|n\rangle, \tag{A2}$$

where the same notations as in [21] are used. The displaced number states (A2) are defined by two numbers: quantum discrete number $n$ and classical continuous parameter $\alpha$ which can be recognized as their size [21]. The states (A2) belong to vector (Hilbert) space with appropriate inner product $\langle n,\alpha|m,\alpha\rangle=\delta_{nm}$ with $\delta_{nm}$ being Kronecker delta [24]. The Hilbert space is determined by the displacement amplitude $\alpha$ with the base states

$$\left\{|n,\alpha\rangle, n=0,1,2,\ldots,\infty\right\}. \tag{A3}$$

If the displacement amplitude is $\alpha=0$, then we deal with Hilbert space of the number states

$$\left\{|n\rangle, n=0,1,2,\ldots,\infty\right\}. \tag{A4}$$

As the set of the states (A3) is complete [21], any displaced number state (A2) from different Hilbert space can be represented in the terms of the base states. So, the number state and their displaced counterparts are related with each other as

$$|l,\alpha\rangle = F\sum_{n=0}^{\infty} c_{\ln}(\alpha)|n\rangle, \tag{A5}$$

where multiplier $F=\exp\left(-|\alpha|^{2}/2\right)$ is introduced and the matrix elements are the following [21]

$$c_{\ln}(\alpha)=\frac{\alpha^{n-l}}{\sqrt{l!}\sqrt{n!}}\sum_{k=0}^{l}(-1)^{k}C_{l}^{k}|\alpha|^{2k}\prod_{k=0}^{l-1}(n-l+k+1), \tag{A6}$$

or the same

$$c_{\ln}(\alpha)=\frac{\alpha^{n-l}}{\sqrt{l!}\sqrt{n!}}\begin{pmatrix} n(n-1)\ldots(n-l+1)-C_{l}^{1}|\alpha|^{2}n(n-1)\ldots(n-l+2)+ \\ C_{l}^{2}|\alpha|^{4}n(n-1)\ldots(n-l+3)+\ldots(-1)^{k}C_{l}^{k}|\alpha|^{2k}\prod_{k=0}^{l-1}(n-l+k-1)+ \\ (-1)^{l}|\alpha|^{2l} \end{pmatrix}, \tag{A7}$$

where $C_{l}^{k}=l!/(k!(l-k)!)$ are the elements of the Bernoulli distribution and number product is

$$\prod_{k=0}^{l-1}(n-l+k+1)=n(n-1)\ldots(n-l+1). \tag{A8}$$

It is worth noting that the reverse transformation $c_{\ln}(-\alpha)$ defines number state through their displaced analogies due to unitary nature of the displacement operator (A1). Consider partial cases with $l=0$ and $l=1$ corresponding to expression of coherent state and displaced single photon in the terms of the number states [21]

$$|0,\alpha\rangle = F\sum_{n=0}^{\infty} c_{0n}(\alpha)|n\rangle, \tag{A9}$$

$$|1,\alpha\rangle = F\sum_{n=0}^{\infty} c_{1n}(\alpha)|n\rangle, \tag{A10}$$

with the matrix elements

$$c_{0n}(\alpha)=\frac{\alpha^n}{\sqrt{n!}}, \tag{A11}$$

$$c_{1n}(\alpha)=\frac{\alpha^{n-1}}{\sqrt{n!}}\left(n-|\alpha|^2\right). \tag{A12}$$

The matrix elements (A11,A12) directly stems from general formulas (A6,A7).

The matrix elements (A6,A7) consist of a common factor proportional to $\alpha^{n-l}$ and a polynomial expression of degree $l$ being the maximum of the degree of its monomial over $|\alpha|^2$ which is enclosed in parentheses. Thus, the term $\alpha^{n-l}$ defines the behavior of the matrix elements under change of the displacement amplitude on opposite in sign that leads to

$$c_{ln}(-\alpha)=(-1)^{n-l}c_{ln}(\alpha). \tag{A13}$$

In particular, we have the following rules for decomposition of the even displaced number states $l=2m$

$$c_{2mn}(-\alpha)=(-1)^n c_{2mn}(\alpha), \tag{A14}$$

and of the odd displaced number states $l=2m+1$

$$c_{2m+1n}(-\alpha)=(-1)^{n-1}c_{2m+1n}(\alpha). \tag{A15}$$

In application to the matrix elements of coherent and displaced single photon states, we have

$$c_{0n}(-\alpha)=(-1)^n c_{0n}(\alpha), \tag{A16}$$

$$c_{1n}(-\alpha)=(-1)^{n-1}c_{1n}(\alpha). \tag{A17}$$

The probability distributions of vacuum and single photon over number states displaced on arbitrary value $\alpha$ are defined by

$$P_{0n}(\alpha)=F^2|c_{0n}(\alpha)|^2=\exp\left(-|\alpha|^2\right)\frac{|\alpha|^{2n}}{n!}, \tag{A18}$$

$$P_{1n}(\alpha)=F^2|c_{10n}(\alpha)|^2=\exp\left(-|\alpha|^2\right)\frac{|\alpha|^{2(n-1)}}{n!}\left(n-|\alpha|^2\right)^2, \tag{A19}$$

respectively. It is possible directly to check the matrix elements $c_{mn}(\alpha)$ (A6,A7) satisfy the normalization condition [21]

$$F^2\sum_{n=0}^{\infty}|c_{mn}(\alpha)|^2=F^2\sum_{n=0}^{\infty}|c_{mn}(-\alpha)|^2=1. \tag{A20}$$

The normalization condition for vacuum $\sum_{n=0}^{\infty}P_{0n}(\alpha)=\sum_{n=0}^{\infty}P_{0n}(-\alpha)=1$ and single photon $\sum_{n=0}^{\infty}P_{1n}(\alpha)=\sum_{n=0}^{\infty}P_{1n}(-\alpha)=1$ can be directly checked using (A11,A12).

## Acknowledgement

The work was supported by Act 211 Government of the Russian Federation, contract № 02.A03.21.0011.

**List of figures**

**Figure 1**
A schematic representation of the quantum teleportation protocol of unknown qubit. Alice and Bob share hybrid entangled state (11). Alice has unknown qubit (10) which she wants to teleport to Bob. To do it she mixes her part of the quantum channel with unknown qubit on HTBS with subsequent measurement in first coherent mode and second teleported mode by commercially achievable APDs in the case of $\alpha << 1$. After that she sends the result of her measurement to Bob and he performs unitary transformations ( $Z$ and $H$ operations) on his part of the quantum channel to end up the protocol. SHC means source of hybrid channel.

**Figure 2**
Dependency of fidelity $Fid$ (Eq. (25)) on the displacement amplitude $\alpha$ and transmittance $t$ for input qubit (10) with amplitudes $a_0 = \sqrt{0.5}$, $a_1 = i\sqrt{0.5}$.

**Figure 3(a-e)**
Three-dimensional plots of probabilities (37) (a) $P_0$, (b) $P_1$, (c) $P_2$ and (d) $P_3$, respectively, in dependency on the displacement amplitude $\alpha$ and absolute value of amplitude $|a_1|$ of unknown qubit (10). Two-dimensional dependencies (e) show the probabilities $P_0$ (curve 1), $P_1$ (curve 2 ) and their sum $P_0 + P_1 \cong 1$ (curve 3 ) for $\alpha = 0.03$.

**Figure 4(a-d)**
Plots of dependencies of probabilities $P_{00}$, (curve 1) $P_{10}$, (curve 2 ) and $P_{20}$(curve 3 ) (Eqs. (44, 45)) on $|a_1|$. The curves correspond to amplitude modulation of unknown qubit (41). The plots are made for the following values of the displacement amplitude (a) $\alpha = 0.06$, (b) $\alpha = 0.1$, (c) $\alpha = 0.2$ and (d) $\alpha = 0.3$.

**Figure 5(a-d)**
Plots of dependencies of probabilities $P_{01}$ (curve 2 ), $P_{11}$ (curve 1 ) and $P_{21}$ (curve 3 ) (Eqs. (49,50)) on $|a_1|$. The curves correspond to another used amplitude modulation of unknown qubit (46). The plots are made for the following values of the displacement amplitude (a) $\alpha = 0.06$, (b) $\alpha = 0.1$, (c) $\alpha = 0.2$ and (d) $\alpha = 0.3$.

**Figure 6**
Optical scheme used for generation of the hybrid channel (11). The same mechanism of interaction of continuous variable and discrete variable states on HTBS as for quantum teleportation protocol (Fig. 1) is used. Additional interaction of auxiliary coherent state with entangled state (53) is employed to complete the formation process of (11).

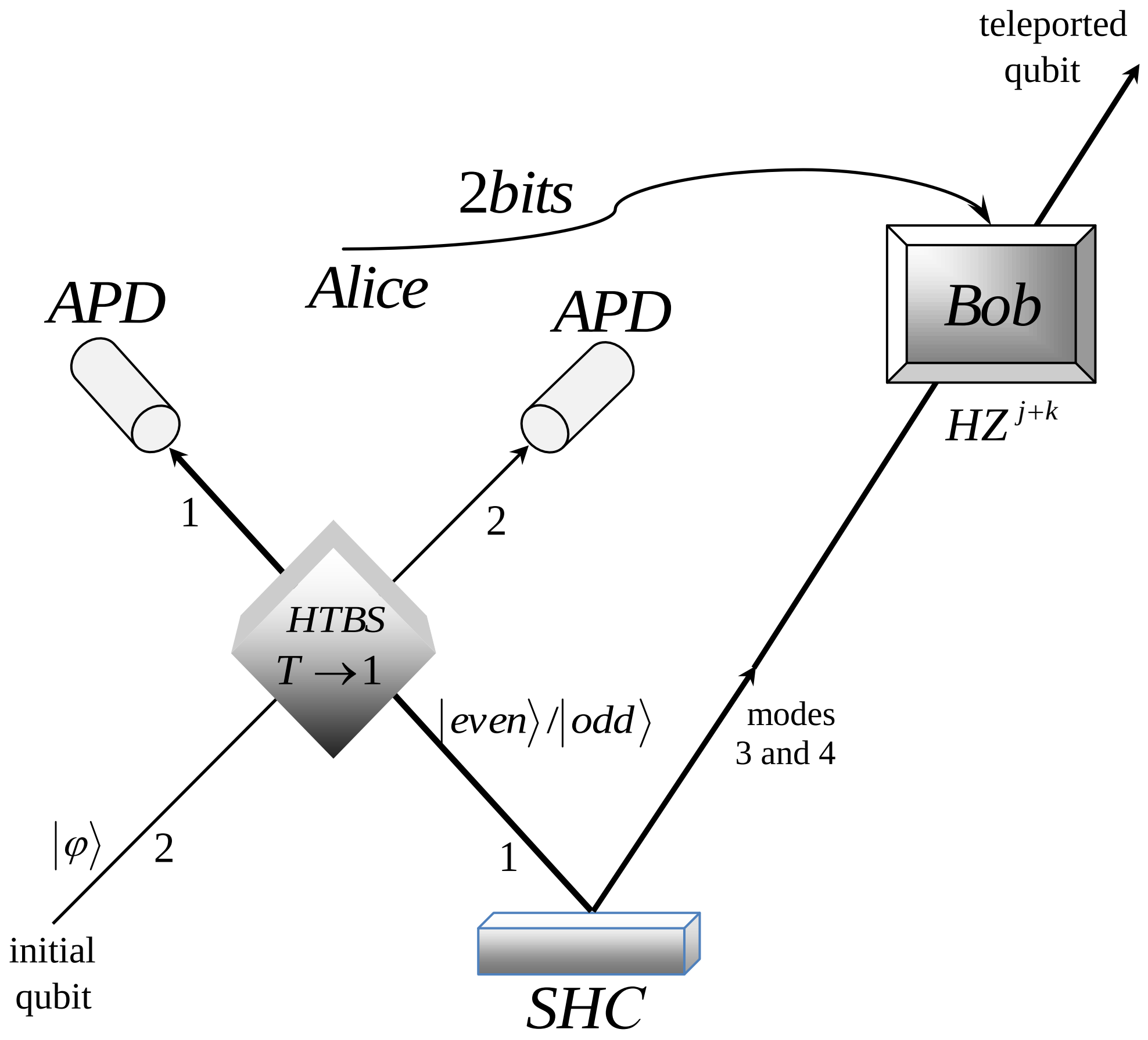

**Figure 1**

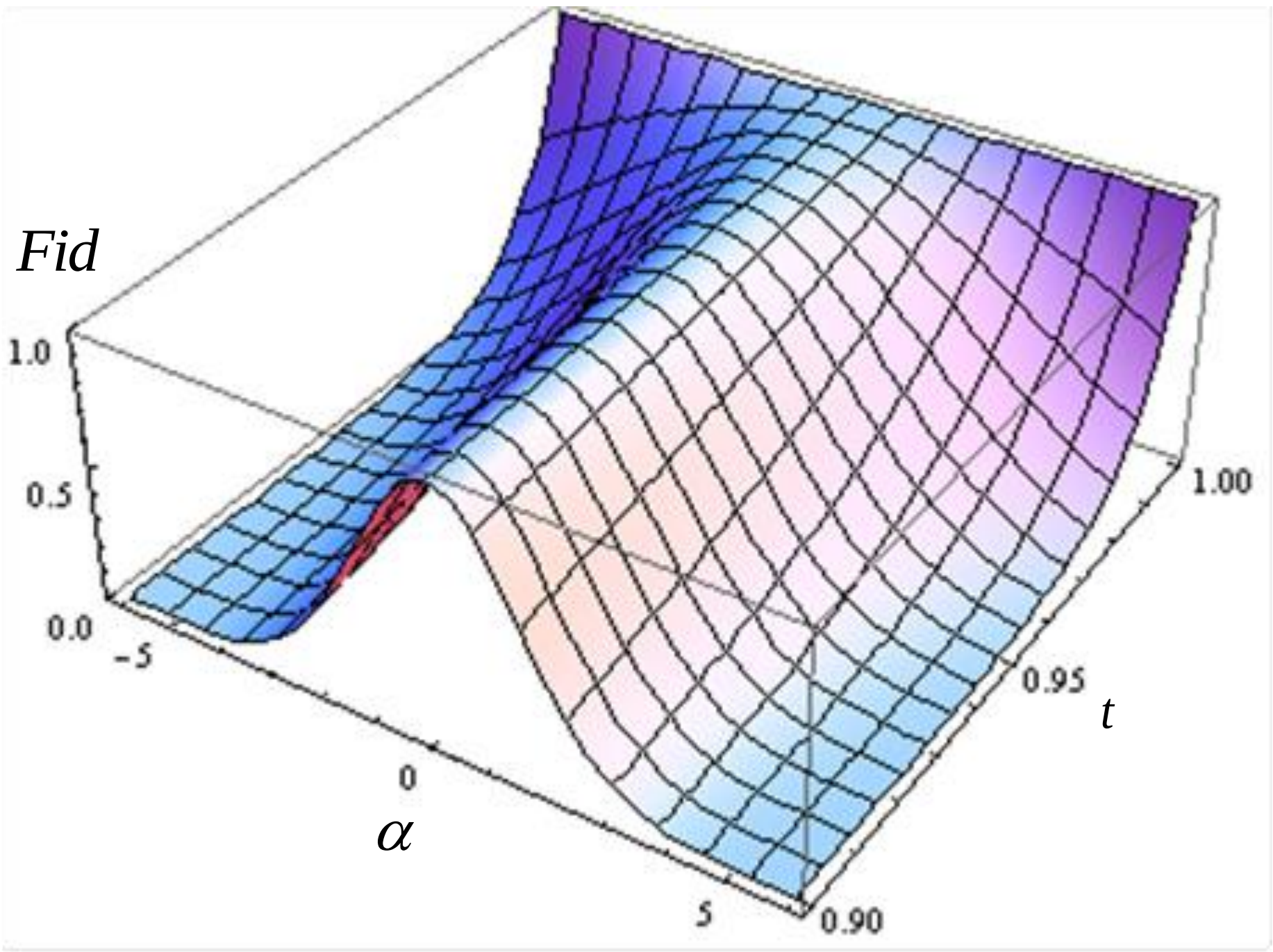

**Figure 2**

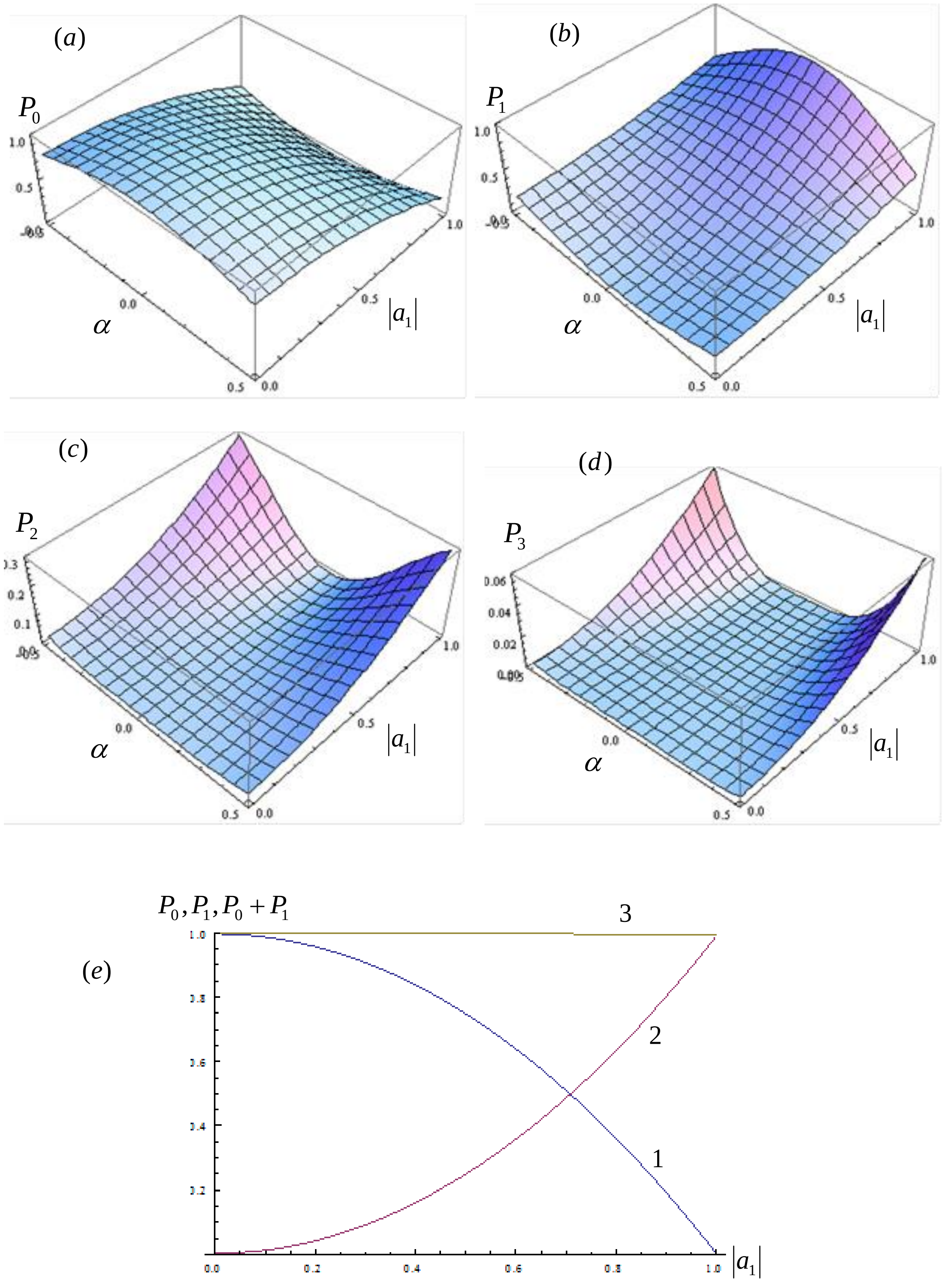

**Figure 3(a-e)**

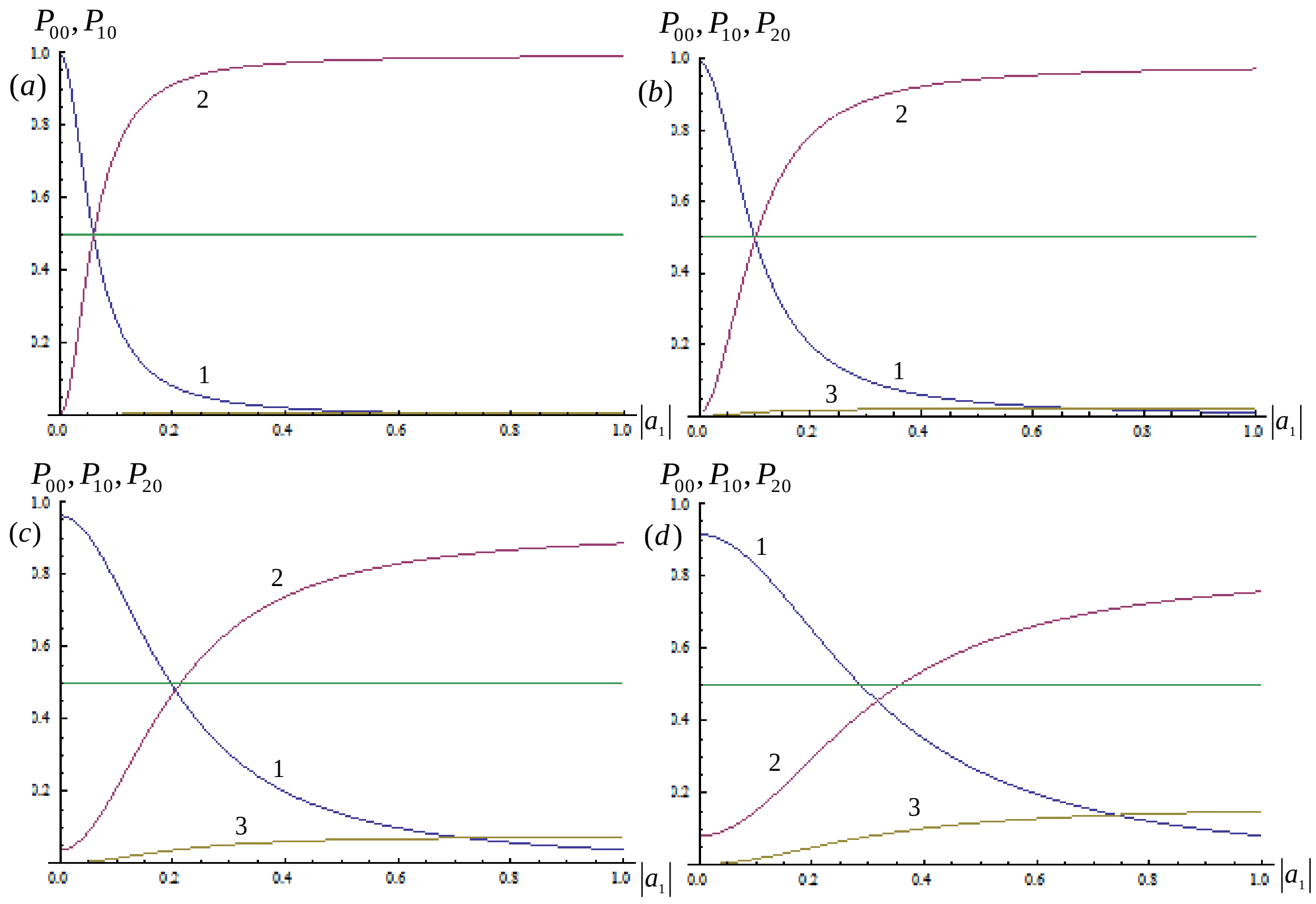

**Figure 4(a-d)**

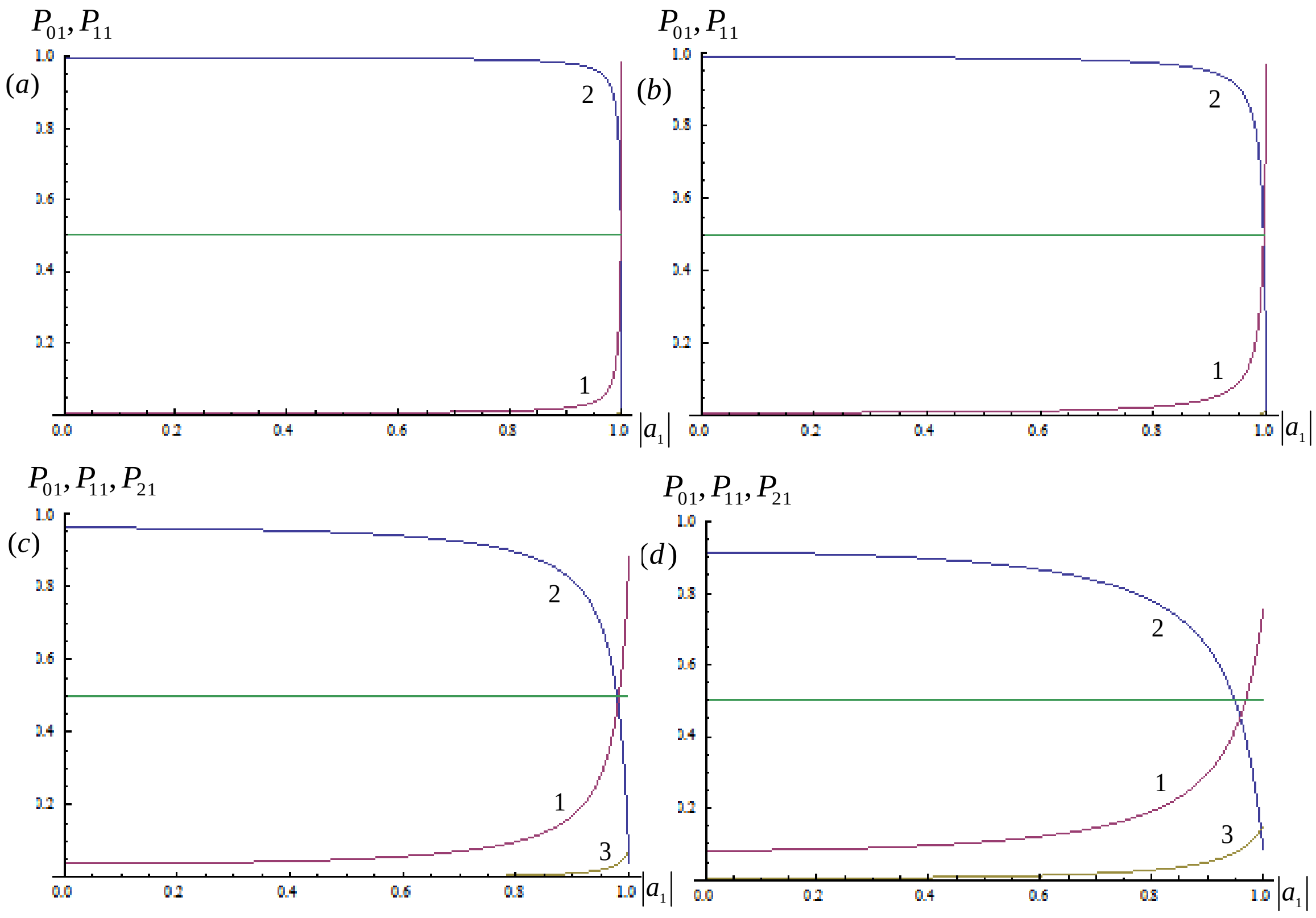

P01, P11
(a)
2
1
|a1|
P01, P11
(b)
2
1
|a1|
P01, P11, P21
(c)
2
1
3
|a1|
P01, P11, P21
(d)
2
1
3
|a1|

**Figure 5(a-d)**

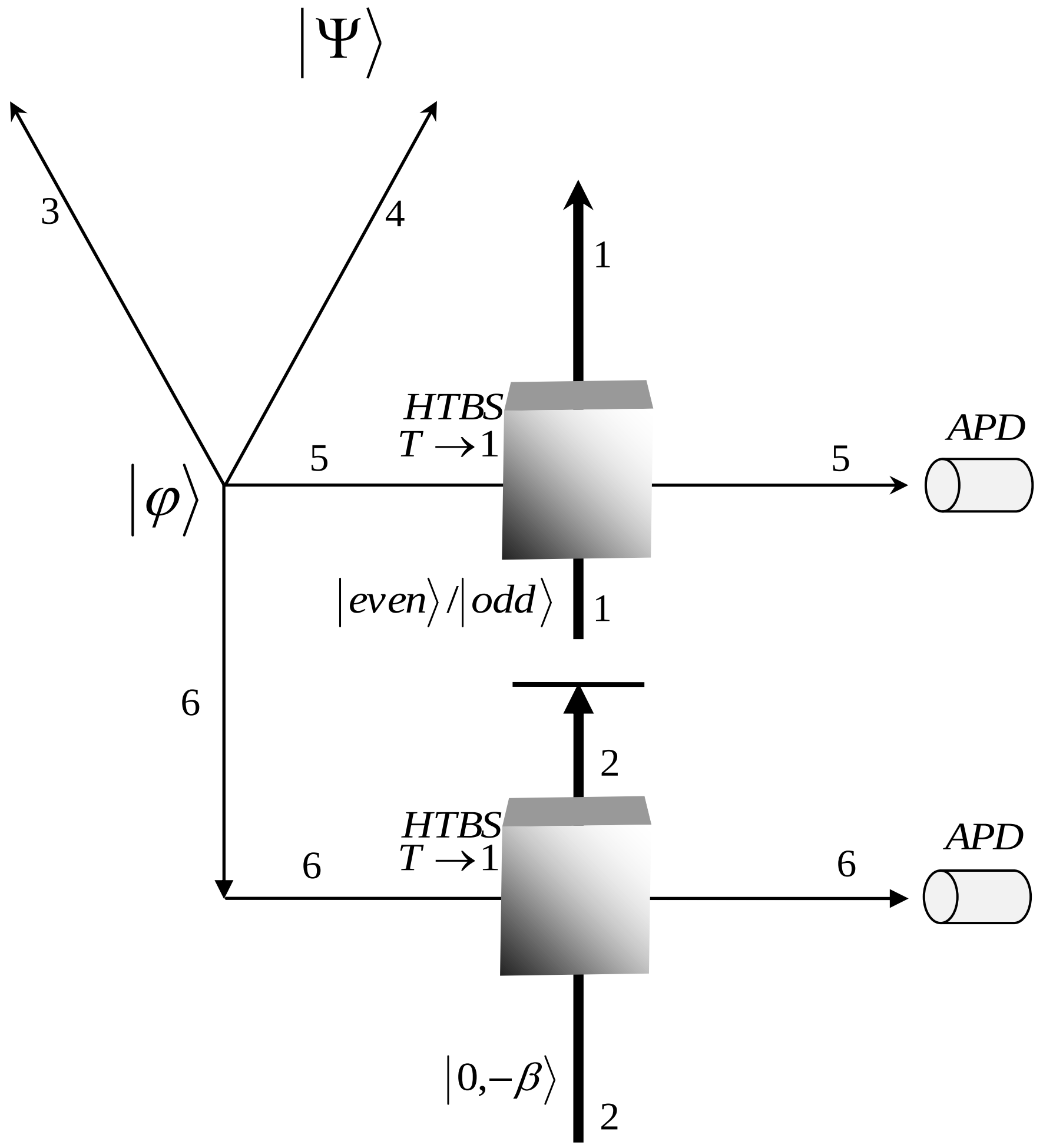

**Figure 6**